\documentclass[aps,amsfonts,nofootinbib]{revtex4}
\newcommand{\bb}{\begin{equation}}
\newcommand{\ee}{\end{equation}}
\newcommand{\eqb}{\begin{eqnarray}}
\newcommand{\eqf}{\end{eqnarray}}

\begin{document}

\title[]{Effects of heavy bosonic excitations on QED vacuum}%
\author{Paola  Arias}
\email{paola.arias@gmail.com}
\affiliation{Departamento de Fisica, Universidad de Santiago de Chile,
  Casilla 307, Santiago Chile}
\author{Horacio Falomir}
\email{falomir@fisica.unlp.edu.ar}
\affiliation{IFLP/CONICET - Departamento de
F\'{\i}sica, Facultad de Ciencias Exactas, Universidad Nacional de la
Plata, C.C. 67, (1900) La Plata, Argentina}
\author{Jorge Gamboa}
\email{jgamboa55@gmail.com}
\affiliation{Departamento de Fisica, Universidad de Santiago de Chile,
  Casilla 307, Santiago Chile}
\author{Fernando M\'endez}
\email{fmendez@fisica.usach.cl}
\affiliation{Departamento de  F\'{\i}sica, Universidad de  Santiago de
  Chile, Casilla 307, Santiago Chile}

\begin{abstract}
We  discuss the contribution  of axion-like  excitations (ALE)  to the
vacuum birrefringence in the limit $m_A \gtrsim \omega$, where $m_A$ is the
mass  of  the excitation  and  $\omega$  the  energy of  test photons
interacting with an external (intense) magnetic field.  The relevance of
this term with respect to the QED contribution depends on
the  ratio $g_A/m_A$  and, from  present bounds  on the  mass  and the
coupling constant $g_A$, we find that in the present low energy
regime , it ranges from $10^{-14}$ to $10^2$ suggesting an interesting
alternative to explore.

\end{abstract}
\maketitle

\section{Introduction}
It is well known that, as a consequence of vacuum polarization effects
in  conventional QED  \cite{HE,Weiss,dittrich,Sch}, the  electromagnetic vacuum
presents a birefringent  behavior \cite{KN,Adler,BI,Bil}. For example,
in the presence of an  intense static background magnetic field, there
are two different refraction  indices depending on the polarization of
the incident electromagnetic wave. This  is in fact a very tiny effect
which could  be measured in  the next few years  \cite{Wipf}.
The  present  bounds  for   this  anisotropy,  followed  from  several
experiments   \cite{BFRT1,BFRT2,PVLAS,PVLAS1},
are still far from the expected value from QED.

\smallskip

According to the original calculations \cite{Adler,BI,adler1,P1,P2,P3,P4,Bil},
conventional QED yields for the difference of the refractive indices
$\Delta n= n_{||}-n_\perp$
\bb
\Delta  n= \frac{\alpha}{45  \pi} \left(\frac{B_0}{E_c}\right)^2  \, ,
\label{t}
\ee
where the \emph{critical  field} $E_c= m_e^2 c^3/\hbar e$,  $B_0$ is the
external (constant  and uniform) magnetic  field and $||$  and $\perp$
refer to  the polarization of  a test wave  with respect to  the plane
determined by the wave vector and the external magnetic field.

Experiments designed to measure such difference are in operation
presently, with  external magnetic fields  ranging between $2$ and
$5$ T (see for instance \cite{BFRT1,BFRT2,PVLAS}). {\sf Then, } one
would expect to measure for the difference of refractive indices
\begin{equation}
    \Delta n \approx 10^{-23} - 10^{-22} \, .   \label{theo}
\end{equation}

As previously mentioned, this anisotropy has not yet been measured,
but  the experimental  results reported  in  \cite{BFRT1,BFRT2, PVLAS}
have set the bound
\bb
\Delta n \leq10^{-19}.
\label{cotas}
\ee

News   are  expected  in  the  near   future  since  interesting
  experiments  on vacuum  birefringence have  been  recently proposed.
  See {\it e.g} \cite{bmv,Wipf}.

\smallskip

On the other hand, if the  interaction of photons  with the magnetic
 field is  also mediated  by additional axion-like  excitations (ALE),
 new  terms will contribute  to the anisotropy through axion-photon conversion
\cite{P1}   and these experiments  can be  used, for  instance, to detect
axions  \cite{mapeza,raffelt}.

In the present paper we will consider this kind of interaction -- an effective
 manifestation  of more fundamental  degrees of  freedom -- in  the low
 energy regime as described by a non-renormalizable Lagrangian.

More  precisely, for  $\varphi$ and $\chi$  a pseudoscalar  and a
scalar  field which respectively couple  to   the  photon  field
invariants ${\cal G}$ and ${\cal F}$ given by
\begin{equation}\label{2}
   \begin{array}{c}
     \mathcal{F}:=   -  \frac{1}{4}  {F}_{\mu   \nu}  F^{\mu   \nu}  =
     \frac{1}{2} \left( \mathbf{E}^2-\mathbf{B}^2
     \right)\,, \\ \\
     \mathcal{G}:= - \frac{1}{4} {\widetilde{F}}_{\mu \nu} F^{\mu \nu}
     = \mathbf{E} \cdot \mathbf{B}\,,
   \end{array}
\end{equation}
where  ${\tilde  F}_{\mu  \nu}  = \frac{1}{2}  \epsilon_{\mu  \nu  \rho
\lambda}F^{\rho \lambda}$, the conventional  QED Lagrangian is replaced by  the
{\it total} Lagrangian
\bb
     \mathcal{L}     =     \mathcal{F}     +     \bar{\psi}     \left[
       \gamma^\mu\left(\imath \partial_\mu - e A_\mu \right) -m_e
    \right] \psi + \mathcal{L}_A + \mathcal{L}_S \, ,
\label{la1}
\ee
where
\begin{eqnarray}
{\cal L}_A &=& \frac{1}{2} \partial_\mu \varphi\, \partial^\mu \varphi -
\frac{1}{2} {m_A}^2 {\varphi}^2 -g_A\,
\varphi\, \mathcal{G}\,,\label{AA}
\\
\mathcal{L}_S &=& \frac{1}{2}  \partial_\mu \chi\, \partial^\mu \chi -
\frac{1}{2} {m_S}^2 {\chi}^2 -g_S\, \chi\, \mathcal{F} \,. \label{S}
\end{eqnarray}

The experiments  should elucidate  which kind of  additional effective
bosonic degrees of freedom are present in this situation, if any.  The purpose
of the present paper is  to discuss the modiffications of (\ref{t}) in
the low  energy limit  $ \omega<<m_a $,  where $\omega$ is  the photon
energy and $m_a$, the mass of the ALE.

We   compute   the   contribution  to   the
electromagnetic   field   effective   Lagrangian   coming   from   the
pseudoscalar field in a gradient approximation, which adds a term to
the    well-known   Heisenberg    -    Euler   effective    Lagrangian
\cite{HE,Weiss,Sch}. Then, we analyze different scenarios for this
extra term, considering  present bounds for $g_A$ and  $m_A$.  We also
show that  the phase shift produced  by the vacuum  birefringence on a
test electromagnetic  due to a scalar ALP differs in sign  with respect to the contribution of a pseudo-
scalar ALE (in agreement with \cite{Ahlers}).

The  paper  is organized  as  follows:  in  Section \ref{Ef-Lag},  the
effective  Lagrangian  piece  coming   from  an  ALE  coupled  to  the
electromagnetic field  is computed  in a gradient expansion.  In the
next section we calculate the refractive indices modified by the presence
of a background magnetic field, in  this low photon energy approximation,
and discuss its relevance for different values of
$m_A$ and $g_A$.
Finally, in Section  \ref {Conclusions} we
give our conclusions and outlook.

\section{The effective Lagrangian}\label{Ef-Lag}
In  this section  we  compute the  analogous  of the  Heisenberg-Euler
effective Lagrangian for the system involving the electromagnetic, the
fermionic and a pseudoscalar fields. In a second part of this section
we will discuss the contribution of a scalar ALE.

As is  well known,  (for a slowly varying electromagnetic field)
the functional integral  over the  fermionic field
leads to the standard Heisenberg-Euler effective Lagrangian,
\cite{HE,Weiss,Sch,Dunne}, {\it i.e}
\begin{equation}\label{3}
   \mathcal{L}_{HE}=   \frac{e^4}{360\,   \pi^{2}  {m_e}^4}   \left\{4
   \mathcal{F}^2 + 7 \mathcal{G}^2 \right\}+
   \cdots \,,
\end{equation}
where the dots represent  higher order terms in $ \mathcal{F}/{E_c}^2$
and $ \mathcal{G}/{E_c}^2$.

Since  the  pseudoscalar is  linearly  coupled  to $\mathcal{G}$,  the
functional integral for this field can also be explicitly evaluated to
get
\begin{equation}\label{4}
    \begin{array}{c}
      \displaystyle
      \int  \mathcal{D}\varphi \,  \exp\left\{ -\frac{i}{2}  \int d^4x
      \left[ \varphi \left( \partial^2 + {m_A}^2 \right)
      \varphi + 2 g_A\,
      \varphi\, \mathcal{G} \right] \right\}=
      \\ \\
      \displaystyle
      =  \left[   {\rm  Det}  \left(  \partial^2   +  {m_A}^2  \right)
      \right]^{-1/2} \exp\left\{ \frac{i}{2}
      \,  g_A^2 \int  d^4  x  \int d^4  y  \, \mathcal{G}(x)\,  K(x,y)\,
      \mathcal{G}(y) \right\}\,,
    \end{array}
\end{equation}
where $K(x,y)=\left( \partial^2 + {m_A}^2 \right)^{-1}(x,y)$.

The exponential in the right  hand side represents a non-local term in
the effective action which, in  the infrared limit we will be interested
in, admits the asymptotic gradient expansion
\begin{equation}
 \frac{i}{2}
      \,   g^2   \int  d^4   x   \int   d^4   y  \,   \mathcal{G}(x)\,
      K(x,y)\                  \mathcal{G}(y)                 =\frac{i
        g_A^2}{2m_A^2}\sum_{n=0}\frac{(-1)^n}{m_A^{2n}}            \int
      d^4x~{\cal G}(x ) ~\partial^{2n}~{\cal G}(x)\,,
\end{equation}
which is justified for $m_A \gg \omega$, where  $\omega$ is an
energy scale  characteristic of  the experiment
under consideration (see Appendix \ref{large-mass}).
Therefore, in this approximation,  the  effective Lagrangian  for the electromagnetic field can be written as the (formally local)
expression
\begin{equation}\label{6}
    \begin{array}{c}
    \displaystyle
      \mathcal{L}_{eff}=   \mathcal{F}(x)   +  \frac{e^4}{360\,   \pi^{2}
      {m_e}^4} \left\{4 \mathcal{F}(x)^2 + 7 \mathcal{
      G}(x)^2 \right\} + \frac{g_A^2  }{2 {m_A}^2}~\sum_{n=0}^\infty\frac{(-1)^n}{m_A^{2n}}{\cal  G}(x )\,
      \partial^{2n}\, {\cal G}(x)\,.
    \end{array}
\end{equation}

\medskip

Let us remark that integrating out the pseudo-scalar field is
equivalent to eliminating it by means of the Euler- Lagrange equation
of motion derived from the quadratic Lagrangian in Eq.\ (\ref{AA}).
In  this respect, one should notice that our
calculation differs from that  of Raffelt and Stodolsky \cite{raffelt}
in  the range of  validity of  the respective  approximations. Indeed,
Ref.\  {\cite{raffelt}}  considers  the  relativistic  limit  for  the
pseudo-scalar,   where   $m_A   \ll   \omega$,   obtaining   non-local
contributions  to  the  equation  of  motion  of  one  of  the  photon
propagation  modes  (see  Eqs.\  (4)-(5) in  \cite{raffelt}).

\smallskip
Similarly, the coupling with a scalar field $\chi(x)$ as in Eq.\ (\ref{S}) adds to the
gradient expansion of the effective Lagrangian in Eq.\ (\ref{6}) the piece
\begin{equation}\label{delta-Leff-escalar}
 \triangle   \mathcal{L}_{eff}   =   \frac{{g_S}^2   }{2   {m_S}^2}
\sum_{n=0}^\infty\frac{(-1)^n}{m_S^{2n}}\,\mathcal{F}(x) \,
      \partial^{2n}\mathcal{F}(x)\,.
\end{equation}

Therefore, we get
\begin{equation}\label{L-eff-total}
    \begin{array}{c}
    \displaystyle
      \mathcal{L}_{eff}
      = \mathcal{F}(x) + \frac{1}{2} \left\{\left[4 \varrho\,\mathcal{F}(x)^2
      +S \sum_{n=0}^\infty\frac{(-1)^n}{m_S^{2n}}\,\mathcal{F}(x) \,
      \partial^{2n}\mathcal{F}(x) \right]+ \right.
      \\ \\
      \displaystyle \left. +
\left[ 7\varrho\, \mathcal{G}(x)^2 + A \sum_{n=0}^\infty\frac{(-1)^n}{m_A^{2n}}{\cal  G}(x )\,
      \partial^{2n}\, {\cal G}(x)\right] \right\}\,,
    \end{array}
\end{equation}
where the parameters $\rho$, $A$ and $S$ are defined as
\begin{equation}\label{A}
    \begin{array}{c}
      \displaystyle
      \varrho:=   \frac{e^4}{180\,  \pi^{2}  {m_e}^4}=   \frac{4  \,
	\alpha^2}{45\, {m_e}^4}\,,
       \\ \\
      \displaystyle
      A:= \frac{g_A^2 }{{m_A}^2}\,,
      \\ \\
      \displaystyle
      S:= \frac{g_S^2 }{{m_S}^2}\,.
    \end{array}
\end{equation}

In the  next section we will calculate  the contribution of these
  terms to  the refractive indices and discuss possible physical
  implications.

\section{Polarization   phenomena   in  electromagnetic   backgrounds}
\label{polarization}
The  polarization  of the  fermionic  vacuum  in  the presence  of  an
electromagnetic background makes it  to act like a birefringent medium
with two  different indices of refraction, depending  on the direction
and  polarization of  the  propagating wave  \cite{BI,Adler}. In  this
Section we evaluate the contributions to the refractive indices due to
the additional coupling with the pseudoscalar and scalar field.

Following \cite{BI}, we will consider the expression in Eq.\ (\ref{L-eff-total})
as the effective Lagrangian for the \emph{total} electromagnetic field, consisting in
the sum of an intense constant uniform background field, $F_{\mu \nu}$, and
a    test    wave     of    low    frequency    (large    wavelength),
$f_{\mu\nu}=\partial_\mu a_\nu - \partial_\nu a_\mu$, and look for the
Euler - Lagrange equations of motion for the last one.

In so doing, it is sufficient to retain the piece quadratic in $f_{\mu
\nu}$,  since the linear  term is  a total  divergence for  a constant
background, and the  higher order terms (cubic and  quartic in $f_{\mu
\nu}$) are suppressed by factors of the order of the ratio between the
intensity of the fluctuation and that of the background.

We get for that piece
\begin{equation}\label{8}
    \mathcal{L}^{(2)}= -\frac{1}{4}\, f_{\mu  \nu}\, M^{\mu\nu\alpha\beta}\,
    f_{\alpha\beta}
\end{equation}
where the operator
\begin{equation}\label{Op-M}
    \begin{array}{c}
      \displaystyle
      M^{\mu\nu\alpha\beta}:= \frac{1}{2}\left[1+(4\varrho+S) \mathcal{F} \right]
    \left(g^{\mu\alpha} g^{\nu\beta} - g^{\mu\beta} g^{\nu\alpha} \right)
    +\frac{1}{2}\left( 7\varrho+A \right) \mathcal{G} \, \epsilon^{\mu\nu\alpha\beta}-
      \\ \\
      \displaystyle
      - \frac{1}{2}\,  F^{\mu\nu} F^{\alpha\beta}\left[4\varrho+ S
      \sum_{n=0}^\infty \left( \frac{- \partial^2}{m_s^2}\right)^n \right]
      - \frac{1}{2}\,  {\widetilde{F}}^{\mu\nu} {\widetilde{F}}^{\alpha\beta}\left[7\varrho+ A
      \sum_{n=0}^\infty \left( \frac{- \partial^2}{m_A^2}\right)^n \right]
    \end{array}
\end{equation}
satisfies $M^{\mu\nu\alpha\beta}= M^{\alpha\beta\mu\nu}=-M^{\nu\mu\alpha\beta}$ and commutes with the derivatives
$\partial_\lambda$ (for a cosntant background field $F_{\mu\nu}$). The corresponding equation of motion for the fluctuation
writes simply as $M^{\mu\nu\alpha\beta}\partial_\nu \partial_\alpha a_\beta=0$.

\smallskip

Let   us  now  look   for  a   solution  of   the  form   $a_\beta(x)  =
\varepsilon_\beta({k})\, e^{- i {k}\cdot {x}} $ which,
replaced in the previous equation, leads to
\begin{equation}\label{9}
    \begin{array}{c}
    \displaystyle
    \frac{1}{2}\left[1+(4\varrho+S) \mathcal{F} \right]
    \left[(k\cdot \varepsilon)k^{\mu}  - k^2 \varepsilon^{\mu} \right]
    +
      \\ \\
      \displaystyle +
      \frac{1}{2} \left(k_\nu F^{\nu\mu}\right) \left(k_\alpha F^{\alpha\beta}\varepsilon_\beta\right)
      \left[4\varrho + S
      \sum_{n=0}^\infty \left( \frac{ k^2}{m_s^2}\right)^n \right]
      +\frac{1}{2} \left(k_\nu {\widetilde{F}}^{\nu\mu}\right) \left(k_\alpha{\widetilde{F}}^{\alpha\beta} \varepsilon_\beta\right)
      \left[7\varrho+ A
      \sum_{n=0}^\infty \left( \frac{k^2}{m_A^2}\right)^n \right]=0\,.
\end{array}
\end{equation}
In the asymptotic large mass expansion we are employing, these series can be summed up to get
\begin{equation}\label{99}
    \begin{array}{c}
    \displaystyle
    \frac{1}{2}\left[1+(4\varrho+S) \mathcal{F} \right]
    \left[(k\cdot \varepsilon)k^{\mu}  - k^2 \varepsilon^{\mu} \right]
    +
      \\ \\
      \displaystyle +
      \frac{1}{2} \left(k_\nu F^{\nu\mu}\right) \left(k_\alpha F^{\alpha\beta}\varepsilon_\beta\right) \left[4\varrho+ S(k)\right]
      +\frac{1}{2} \left(k_\nu {\widetilde{F}}^{\nu\mu}\right) \left(k_\alpha{\widetilde{F}}^{\alpha\beta} \varepsilon_\beta\right)
      \left[7\varrho+ A(k) \right]=0\,,
\end{array}
\end{equation}
where we have called
\begin{equation}\label{AyS}
    A(k):= \frac{A}{1-k^2/m_A^2}\,,\qquad
    S(k):= \frac{S}{1-k^2/m_S^2}\,.
\end{equation}

Eq.\ (\ref{99}) implies that the polarization vector must have the form
\begin{equation}\label{10}
    { \varepsilon^\nu({k}) = \xi_0 k^\nu + \xi_1 k_\mu F^{\mu\nu}
    +\xi_2 k_\mu \widetilde{F}^{\mu\nu}}\,.
\end{equation}
where  $\xi_0$  represents  a  gauge  transformation and  can  not  be
determined from those (gauge invariant) equations.

Taking into account that
 \begin{equation}\label{11}
    \begin{array}{l}
      k^\mu F_{\mu\nu}  \widetilde{F}^{\nu \lambda} k_\lambda  = {k}^2
      \mathcal{G}\,,
      \\ \\
      k^\mu \widetilde{F}_{\mu\nu} \widetilde{F}^{\nu \lambda} k_\lambda =
       k^\mu  F_{\mu\nu} {F}^{\nu \lambda}  k_\lambda -  2 \mathcal{F}
      \,{k}^2 \,,
    \end{array}
\end{equation}
it is straightforward to show that the propagation normal modes satisfy
\begin{widetext}
\begin{equation}\label{12}
    \left(
      \begin{array}{cc}
        { \left[1 + (4\varrho+S) {\mathcal{F}} \right] {k}^2 + \left[4\varrho+S(k)\right] { k\cdot
         F\cdot F\cdot k }}
         &
        { \left[ 4\varrho+S(k) \right]    \mathcal{G} \,{k}^2 }
        \\ \\
       { \left[ 7\varrho + {A(k)}\right]  \mathcal{G} \,{k}^2}   &
       { \left[1  -\left(10\varrho+2{A(k)}-S\right) \mathcal{F} \right]  {k}^2 + \left[7\varrho+ {A}(k)\right]
          {k\cdot {F}\cdot {F}\cdot k}}\\
      \end{array}
    \right)
    \left(
      \begin{array}{c}
        \xi_1
        \\ \\
        \xi_2 \\
      \end{array}
    \right) =0\,.
\end{equation}
\end{widetext}

Here
\begin{equation}\label{13}
    \begin{array}{c}
    \displaystyle
      { k\cdot F\cdot  F \cdot k }:= k^\mu  F_{\mu\nu} F^{\nu \lambda}
      k_\lambda =
      \\ \\
      \displaystyle =
      {\omega}^2 \mathbf{E}^2  + \mathbf{k}^2 \mathbf{B}^2  - 2 \omega
      \mathbf{k}\cdot \mathbf{E}\times \mathbf{B}
-\left(  \mathbf{k}\cdot  \mathbf{E} \right)^2-\left(  \mathbf{k}\cdot
      \mathbf{B} \right)^2 \,,
    \end{array}
\end{equation}
where  $\omega=k_0$,   and  $\mathbf{E}$  and   $\mathbf{B}$  are  the
background electric and magnetic fields, respectively.

Nontrivial solutions of Eq.\ (\ref{12}) require the determinant of the
matrix  on the  left  hand side  to  vanish.  If,  for simplicity,  we
consider the case in which { $\mathcal{G}=0$}\ \ (which corresponds to
$\mathbf{E}  \perp  \mathbf{B}$, or  pure  magnetic  or pure  electric
background field),  the non-diagonal  elements of this  matrix vanish,
and the normal  modes are determined by setting  the diagonal elements
equal to zero.

For  definiteness,  let  us  consider  the case  of  a  pure  magnetic
background    field:    $F_{12}=B_0=-F_{21}$    (    $\mathbf{B}_0=B_0\,
\mathbf{e}_3$) and the other components  equal to zero. We get for the
two normal modes
\begin{equation}\label{14}
   \begin{array}{l}
     \left[1-\frac{1}{2} (4\varrho+S) {B_0}^2\right]   \left({\omega_1}^2
 -{\mathbf{k}}^2\right) + \left[4\varrho+S(k)\right]  {B_0}^2\, {\mathbf{k}_\perp}^2=0\,,
 \\ \\
      \left[1+\frac{1}{2}(10\varrho+2{A}(k)-S)   {B_0}^2\right]    \left({\omega_2}^2
 -{\mathbf{k}}^2\right) + \left[7\varrho+{A}(k)\right]
      {B_0}^2\, {\mathbf{k}_\perp}^2 =0\,,
   \end{array}
\end{equation}
where  $k=(\omega,\mathbf{k})$ and ${\mathbf{k}_\perp}$   is   the  component   of   $\mathbf{k}$
perpendicular  to  the  magnetic  field. This  implies  the
dispersion relations implicitly given by the following equations
\begin{equation}
\label{15}
   \begin{array}{l}
   \displaystyle
   {\omega_1}^2 =  {{\mathbf{k}}}^2 -
     \left(\frac{\left(4  \varrho + S(k)  \right) {B_0}^2}{{1-\frac{1}{2}\left(
     4  \varrho+ S \right)  {B_0}^2}}\right)
     {\mathbf{k}_\perp}^2
    \,,
     \\ \\
     \displaystyle
     {\omega_2}^2 =  {{\mathbf{k}}}^2 -
     \left(\frac{\left(7\varrho +{A(k)}\right) {B_0}^2}{{1+\frac{1}{2}\left(10\varrho+2{A(k)}-S\right)
         {B_0}^2}} \right) {\mathbf{k}_\perp}^2
    \,.
   \end{array}
\end{equation}

 Notice that, in these  equations, ${A(k)}$ and ${S(k)}$ also depend on the normal frequencies $\omega_1$ and $\omega_2$.

These equations can be solved in a large mass expansion (consistent with the gradient expansion we have employed) to get,
for example,
\begin{equation}\label{omegas}
    \begin{array}{c}
    \displaystyle
      {\omega_1}^2=
      {\mathbf{k}}^2\left\{ \left[1-(4 \rho + S) B_0^2 \sin^2 \theta
      -\frac{1}{2}(4 \rho + S)^2 B_0^4 \sin^2 \theta\right]+
      \right. \\ \\ \displaystyle \left. +
      \frac{\mathbf{k}^2}{m^2}
      \left[ {  (4 \rho + S) S\, B_0^4 \sin^4 \theta } \right]
      +O\left[\left(\frac{\mathbf{k}^2}{m^2}\right)^2\right]\right\}\,,
      \\ \\ \displaystyle
      {\omega_2}^2={\mathbf{k}}^2 \left\{ \left[1-
      (7 \rho+A) B_0^2 \sin^2 \theta
      +\frac{1}{2} (7 \rho+A)(10 \varrho+2A-S) B_0^4 \sin^2 \theta
      \right]+
      \right. \\ \\ \displaystyle \left. +
      \frac{{\mathbf{k}}^2}{m^2}\left[
        { (7 \rho +A) A\,B_0^4 \sin^4 \theta} \right]+
        O\left[\left(\frac{{\mathbf{k}}^2}{m^2}\right)^2\right] \right\}\,,
    \end{array}
\end{equation}
 where, in the right hand sides, we have retained just the first terms in $\varrho {B_0}^2$, $A {B_0}^2$, and $S {B_0}^2$,
 parameters whose values will be consider comparable for the time being. The angle $\theta$ is defined by
 $\cos \theta= \mathbf{k}\cdot \mathbf{B}_0 / |\mathbf{k}| B_0$.

Let us now  introduce  the   refractive  index  as ${n}(\omega):= \left| {\mathbf{n}}(\omega) \right|$, where
${\mathbf{n}}(\omega):={\mathbf{k}}/\omega$, and discuss the anisotropy induced by the non-linear terms in the effective
Lagrangian. From Eq.\ (\ref{15}) we get for the indices and polarization vectors
\begin{widetext}
\begin{equation}\label{16}
   \begin{array}{l}
   \displaystyle
   {n_1(\omega)}^2 = 1+(4 \rho +S )B_0^2 \sin ^2\theta
   -\frac{1}{2}(4 \rho +S )^2B_0^4 (\cos 2\theta -2) \sin ^2\theta
   -
   \\ \\ \displaystyle
   \hfill- \frac{\omega^2}{m^2} \left[
   { (4\rho +S) S\,  B_0^4 \sin^4 \theta }
   \right]+O\left[\left(\frac{\omega}{m}\right)^4\right]
   \,,
   \\ \\ \displaystyle
     \varepsilon_1= \mathbf{n}_1 \times {
     {\mathbf{B}_0}}\,,
     \\ \\
     \displaystyle
     {n_2(\omega)}^2=1+(7 \rho+A) B_0^2 \sin ^2 \theta
     -\frac{1}{2} (7 \rho+A) \left[3 \varrho+A-S +(7 \varrho +A) \cos 2\theta\right] B_0^4 \sin ^2 \theta-
     \\ \\ \displaystyle
     \hfill-\frac{\omega^2}{m^2} \left[
     {(7\rho+A )A B_0^4 \sin^4 \theta } \right]+
     O\left[\left(\frac{\omega}{m}\right)^4\right]\,,
     \\ \\ \displaystyle
    \varepsilon_2= {{\mathbf{B}_0}} -
     \mathbf{n}_2
     \left( \mathbf{n}_2
     \cdot {{\mathbf{B}_0}} \right)\,,
   \end{array}
\end{equation}
\end{widetext}
for the \emph{transverse} and \emph{parallel} polarizations respectively,
where the terms   {{transverse}}  and   {{parallel}}   refer  to   the
orientation  of the  polarization  vector with  respect  to the  plane
determined  by the  vectors $\mathbf{k}$  and  ${{\mathbf{B}_0}}$ (The
gauge  parameter  $\xi_0$   has  been  chosen  in  such   a  way  that
${\varepsilon_{1,2}}^0=0$).   Notice   that the
right hand sides of  Eq.\   (\ref{16})   with
$A,S\rightarrow 0$ coincide with the result in \cite{BI}.  Moreover, it
agrees with the result by Raffelt and Stodolsky in \cite{raffelt} (see
also  \cite{china,china1}) in the  sense that, in this case,  it is  only the  parallel mode
which is  affected by the presence of  the pseudoscalar ALE (even though  it is the
relativistic small  pseudo-scalar  mass  - limit  which  has  been
considered in that reference).

The difference  in refractive indices induced by  the background field
is maximal  at $\theta = \pi/2$  ($\mathbf{k} \perp {{\mathbf{B}_0}}$)
and    vanishing    for    $\theta    =0$    ($\mathbf{k}    \parallel
{{\mathbf{B}_0}}$).   Moreover,   the  polarization  vector   for  the
parallel mode  is not  perpendicular to $\mathbf{k}$:  { $\mathbf{n}_2
\cdot  \mathbf{\varepsilon}_2 = (\mathbf{n}_2  \cdot {{\mathbf{B}_0}})
(1-{n_2}^2) \neq 0$}.

Notice that the first non-vanishing contribution proportional to $\omega^2/m_{A,S}^2$ is $O\left(\varrho^2 B_0^4\right)$ which,
for typical values of $B_0\sim 1$T, turn out to be of the order $10^{-50}$. Therefore, for practical purposes, it is  enough to
preserve only  the zero  order term  in  the large mass expansion, retaining in it only the first order in $\varrho B_0^2$,
$A B_0^2$ or $S B_0^2$.

Notice that, up to this order of approximation, the presence of the pseudoscalar field reflects in
the refractive  index of  the parallel mode,  while the scalar field  has no
effect on it  and affects only the refractive  index of the transverse
mode.

\smallskip

Different  values for the  refractive indices of the normal modes imply birefringence.
Indeed,  a  linearly  polarized  electromagnetic  wave  of  wavelength
$\lambda= 2\pi/\omega$ which,  for example, propagates perpendicularly
to  the uniform magnetic  field $\mathbf{B}_0$  along a  distance $L$,
turns into an  elliptically polarized wave with a  phase shift between the two rays given by
\begin{equation}\label{phi-escalar}
     \phi = 2 \pi  (n_2-n_1) \frac L \lambda \simeq
     \pi (3\varrho+A-S)   {{{B}_0}}^2\, \frac L \lambda \,.
\end{equation}

Therefore, the sign  of the contribution to the phase shift depends on  what kind of bosonic excitation
realizes this low-energy effective  coupling of fundamental degrees of
freedom with the electromagnetic field.  In a first approach, we could
estimate  $S  \approx   A$.  Then,  the  sign  of   $\phi$ would allow  to
discriminate among  the presence of  a scalar or a  pseudoscalar ALE
coupled to  the electromagnetic field.  This is in agreement  with the
analysis in \cite{Ahlers}.

Indeed, in terms of the refraction index arising just from QED, $ \Delta
n_{QED}$ in  (\ref{t}), we have  for the difference of refractive indices
\begin{eqnarray}
\Delta n=\Delta
n_{\mbox{\tiny{QED}}}\left(1+\frac{A-S}{3\varrho}\right)\, ,\label{ult}
\end{eqnarray}

Let us consider  the relative weight of the contributions  from the
fermionic and the pseudoscalar fields. This   is simply given by the
quotient
\begin{equation}\label{18}
    \frac{A}{3\varrho}= \frac{15\, {m_e}^4 {g_A}^2 }{4 \, \alpha^2 \, {m_A}^2}
    =4.4\times10^{27} (\mbox{eV})^4 \times  \left(\frac{g_A}{m_A}\right)^2
\,.
\end{equation}

Bounds for $g_A$ and $m_A$  for the axion-photon coupling can be found in
\cite{pdg}. For example, from the optical experiments reported in
 \cite{BFRT1,BFRT2,PVLAS,gammev},  we can take
$m_A \sim 10^{-3}$ eV and $g_A \sim 3\times 10^{-16}$ eV$^{-1}$, leading to
$$
\frac{A}{3\varrho}\sim 4\times 10^2
$$
for which $\triangle n$ is two orders of magnitud below the bound (\ref{cotas}) and two
orders of magnitud grater than $\Delta n_{\mbox{\tiny{QED}}}$.

For    different     bounds, as for invisible    axions     with    larger
masses \cite{morales,avignon}, $m_a\sim 1$ keV, $g_A\sim
3 \times 10^{-18}$ eV$^{-1}$,
and one has
$$
\frac{A}{3\varrho}\sim 4\times 10^{-14}\,.
$$

The weight of this contribution is, in this case,  negligible with respect to that of
QED.

But  let us  stress  that, strictly speaking, our  results  are valid  in the  large-mass
approximation,  where $m_{A,S}$ is  larger that  the photon  energy in the probe ray. Since   the  frequency  of the laser
employed in the experiments \cite{BFRT1,BFRT2,PVLAS,Adler2008gk} is around 1eV,
our results suggest  to look  for  ALE with  masses  $m_A \gtrsim  1$eV,  which
corresponds  to  a  much  heavier excitation  than  the  conventional
axion-like particles considered previously in this context.

\section{Discussion and Conclusions} \label{Conclusions}

In this  paper we  have speculated about  the role a  heavy axion-like
excitation can play in  the effective electromagnetic phenomena at low
energies.  In   a  large   mass  approximation,  we   have  considered
excitations of both pseudoscalar  and scalar nature, and studied their
contributions to the vacuum birefringence phenomenon.

Even though the effects of a pseudo-scalar axion-like particle on the electromagnetic
vacuum  birefringence and dichroism  properties have  been extensively
considered in the  literature, the novelty of our  research resides on
the  inclusion of  heavy bosonic  excitations in  a  gradient
expansion of the effective Lagrangian (leading to a simple large-mass
approach). The  contribution  of   this  term   to  the   standard  vacuum
  birrefringence predicted  by QED,  depends on the  ratio $g_A/m_A$,
  and it can be numerically  estimated by using present bounds of both
  parameters, arising from  experiments of axion-photon couplings, for
  instance.  We  have  found  that, according to these  bounds,  this
  contribution  could  be  non-negligible.  In any  case,  the  bound
  (\ref{cotas}) is not saturated.

But, strictly  speaking, since the optical experiments  have been done
employing lasers  of frequencies $\omega \approx  1$\,eV, the validity
of our approximation  requires that $m_A \gtrsim 1$eV,  at least three
orders  of magnitude grater  than the  masses of  axion-like particles
previously  considered in this  context. This  gives for  the coupling
constant
 \begin{equation}\label{gA}
    g_A > {10}^{-12} (eV)^{-1}\, .
 \end{equation}

 \smallskip

On the other hand, following \cite{Adler2008gk} we can estimate the photon-ALE conversion probability as
\begin{equation}\label{probability}
    P_{\gamma \rightarrow \mbox{\tiny{ALE}}}
    =\frac{{g_A}^4\,{B_0}^4}{8\,q\,{m_A}^4}\log\left(\frac{2\,q\,{m_A}^4}{{g_A}^4\,{B_0}^4}
    \right)\,,
\end{equation}
where $q$ is  the quality factor of the  laser source ($q=\Delta/m_A$,
with $\Delta$ the laser bandwidth).

Taking,   for  example,   the  data   from  the   LULI-BOSONS  project
\cite{BOSONS}, $q\sim 10^{-5}$ and  $B_0\sim 2\times 10^2$ eV $^2$, we
get a photon-boson conversion  probability of the order of $10^{-34}$,
with an extremely small rotation angle of the polarization plane. This
is consistent with the non-observation of vacuum dichroism.

Although  this probability  is  similar  to the  one  obtained in  the
standard  calculation  in   axion  physics  \cite{Kuster:2008zz},  the
philosophy of  our approach is  quite different and could  justify new
experimental   efforts,   eventually   beyond  the   present   optical
experiments.

\vspace{0.3 cm}

\noindent\underline{Acknowledgements}:  We thank Fidel  Schaposnik for
many  useful  discussions.  This   work  was  partially  supported  by
FONDECYT-Chile and CONICYT under grants 1050114, 1060079 and 21050196,
and by PIP 6160 - CONICET, and UNLP grant 11/X492, Argentina and DICYT (USACH).


\appendix

\section{Large mass approximation}
\label{large-mass}

As pointed out in Section \ref{polarization},
in  order to  determine  the effects  on the  refractive
indices,  it  is sufficient  to  retain  the  piece of  the  effective
Lagrangian  quadratic in  the fluctuating  electromagnetic  field.

For the case  of interest, of a large  constant and uniform background
magnetic field plus a (laser) test wave of frequency $\omega$, we have
\begin{equation}\label{a1}
    \mathcal{G}=  \mathbf{B}  \cdot  \mathbf{E} \simeq  {\mathbf{B}_0}
    \cdot \mathcal{E} + \cdots\,,
\end{equation}
where $\mathcal{E}= \varepsilon \,  e^{- \imath (\omega t - \mathbf{k}
\cdot  \mathbf{x} )}$ is  the electric  field of  the test  wave (with
$\varepsilon$ the polarization  vector) and the dots stand  for a term
quadratic in the fluctuating field.

In terms of the Fourier transform of $\mathcal{G}(x)$,
\begin{equation}\label{a2}
   \widetilde{ \mathcal{G}}(q) = \int  d^4 x \ e^{\imath (q-k)\cdot x}
   \, \mathbf{B}_0 \cdot \varepsilon \,,
\end{equation}
the argument of the exponential factor  in the right hand side of Eq.\
(\ref{4}) reduces to
\begin{equation}\label{a3}
    \frac{1}{2}      \int      \frac{d^4      q}{(2     \pi)^4}      \
    \frac{\widetilde{\mathcal{G}}^\ast(q) \,
    \widetilde{\mathcal{G}}(q)}{-q^2 +
    {m_A}^2}=
    \int d^4 x\ \frac{ \left| \mathbf{B}_0 \cdot \varepsilon \right|^2
    }{-k^2 + {m_A}^2}
    =  \frac{1}{{m_A}^2}  \int  d^4 x\,  {\mathcal{G}(x)}^2  \left[1+O
    \left( \frac{\omega^2}{{m_A}^2} \right) \right]\,,
\end{equation}
showing  that our gradient  approximation applies when $m_{A,S}$ is grater
that the energy of the test (laser) photons.



\begin{thebibliography}{99}
\bibitem{HE} W.\ Heisenberg and H.\  Euler, {\it Z.\ Phys.}\ {\bf 98},
714 (1936).
\bibitem{Weiss}{V.\ Weisskopf, \emph{The electrodynamics of the vacuum
based  on the  quantum theory  of electrons},  English  translation in
Early  Quantum Electrodynamics:  A  source book,  A.I.\ Miller  Edt.\,
Cambridge  University Press (1994).}
\bibitem{dittrich} W. Dittrich  and H.  Gies, {\it
Probing the Quantum Vacuum:  Perturbative Effective Action Approach in
Quantum  Electrodynamics  and Its  Applications},  Springer Tracts  in
Modern Physics 2000.
\bibitem{Sch}  J.\  Schwinger,  {\it  Phys.\  Rev.}  \textbf{82},  664
(1951).
\bibitem{KN} J.\ Klein and  B.\ Nigam, {\it Phys.\ Rev.} \textbf{135},
1279 (1964).
\bibitem{Adler} S.  L. Adler,  {\it Ann. Phys.}  (N.Y.) {\bf  67}, 599
(1971).
\bibitem{BI}E.\ Brezin and C.\ Itzykson, {\it Phys. Rev.} \textbf{D3},
618 (1971).
\bibitem{Bil}  Z.  Bialynicka-Birula  and I.   Bialynicka-Birula, {\it
Phys. Rev.} {\bf D2}, 2341 (1970).
\bibitem{Wipf} T.   Heinzl, B.   Liesfeld, K.  Amthor,  H.  Schwoerer,
R. Sauerbrey and  A. Wipf, {\it Optics Communications}  {\bf 267}, 318
(2006).
\bibitem{BFRT1}  Y.   Semertzidis  et  al [BFRT  Collaboration],  {\it
Phys. Rev. Lett.} {\bf 64}, 2988 (1990).
\bibitem{BFRT2}   R.   Cameron  et   al  [BFRT   Collaboration],  {\it
  Phys. Rev.} {\bf D47}, 3707 (1993).
\bibitem{PVLAS}  E.   Zavattini et  al.   [PVLAS collaboration],  {\it
  Phys. Rev.  Lett.}  {\bf 96}, 110406 (2006).
  \bibitem{PVLAS1} E.  Zavattini, et al. [
  PVLAS collaboration], {\it Phys.  Rev.}  D {\bf 77}, 032006 (2008).
\bibitem{adler1} S. L. Adler {\it J. Phys.} {\bf A40}, F143 (2007).
\bibitem{P1} P. Sikivie, {\it Phys. Rev. Lett.} {\bf 51}, 1415 (1983).
\bibitem{P2} L. F. Abbot and P. Sikivie,  {\it Phys. Lett.}  {\bf B120}, 133 (1983).
\bibitem{P3}  P. Sikivie,  {\it Phys. Rev. Lett.} {\bf 48}, 1156 (1982). 
\bibitem{P4} D. A. Dicus, E. W. Kolb, V. L. Teplitz, R. V. Wagoner,  {\it Phys. Rev.} { \bf D18}, 1829 (1978).
\bibitem{bmv}   R.   Battesti   et   al  [BMV   collaboration],   {\it
Eur. Phys. J.} {\bf D46}, 323 (2008).
\bibitem{mapeza}L. Maiani, R. Petronzio and E. Zavattini, {\it Phys. Lett.} {\bf B175}, 359 (1986).
\bibitem{raffelt} G. Raffelt and L.  Stodolsky, {\it Phys. Rev. } {\bf
D37}, 1237 (1988).
\bibitem{china}W. -Y. Tsai and T.  Erber, {\it Phys. Rev. } {\bf D12},
1132  (1975).
\bibitem{china1} R.  Novick,  M.  C.   Weisskopf, J.  R.  P.  Angel  and
P. G. Sutherland, {\it Astrophys. J.} {\bf 215}, L117 (1977).

\bibitem{Ahlers} M. Ahlers, H. Gies,  J. Jaeckel and A. Ringwald, {\it
Phys.\ Rev.}\ \textbf{D75}, 035011 (2007).
\bibitem{Dunne}{G.   V.    Dunne,   \emph{Heisenberg-Euler   Effective
Lagrangians : Basics and Extensions}, arXiv:hep-th/ 0406216v1.}
\bibitem{pdg}C. Amsler et al. (Particle Data Group),
{\it Physics Letters } {\bf B667}, 1 (2008)
\bibitem{gammev} Aaron S.. Chou et al  [GammeV (T-969) Collaboration]
{\it Phys. Rev. Lett.} {\bf 100}, 080402 (2008).
\bibitem{morales}A. Morales et al. [COSME collaboration] {\it Astropart. Phys. } {\bf
16},  325 (2002).
\bibitem{avignon}
F.T. Avignone, III et al. [SOLAX Collaboration], {\it Phys. Rev. Lett. } {\bf 81} , 5068 (1998).
\bibitem{Adler2008gk}    S.~L.~Adler,    J.~Gamboa,   F.~Mendez    and
  J.~Lopez-Sarrion, {\it Ann.  of Phys.}  {\bf 323}, 2851 (2008).
\bibitem{BOSONS}         See         the         LULI         homepage
http://www.luli.polytechnique.fr/pages/facilites.htm.
\bibitem{Kuster:2008zz}   M.~Kuster,    G.~Raffelt   and   B.~Beltran, 
{\it  Prepared  for  Joint  ILIAS-CAST-CERN Axion  Training  at  CERN,
Geneva, Switzerland, 30 Nov 2 Dec 2005}.


\end{thebibliography}
\end{document}